\documentclass[onecolumn]{revtex4-2}
\usepackage{graphicx}
\usepackage{amsmath}
\usepackage{amsfonts}
\usepackage{amssymb}
\usepackage{epsfig}
\usepackage{color}
\usepackage{dsfont}

\newcommand{\numberset}{\mathbb}

\newcommand{\R}{\numberset{R}}

\newcommand{\ket}[1]{\vert#1\rangle}
\newcommand{\bra}[1]{\langle#1\vert}
\newcommand{\Tr}[1]{\text{Tr}\left\{#1\right\}}

\begin{document}

\title{A note on the time-reversal symmetry for the quasiprobability distributions of work}
\author{Gianluca Francica}
\address{Dipartimento di Fisica e Astronomia e Sezione INFN, Università di Padova, via Marzolo 8, 35131 Padova, Italy}

\begin{abstract}
In this short note we discuss the time-reversal of a quasiprobability distribution of work.
\end{abstract}

\maketitle

\section{Introduction}
For quantum systems there are several proposals to describe the work. Among these, representations in terms of quasiprobability distributions of work have received some attention.
Here, we aim to  briefly discuss the time reversal of certain quasiprobability distributions~\cite{Francica22,Francica222}.


\section{Quasiprobabilities}
We start to recall our notion of quasiprobability~\cite{Francica222}.
In general, events are represented as effects, which are the positive operators which can occur in the range of a positive operator valued measurement, i.e., an effect is a Hermitian operator $E$ acting on the Hilbert space $\mathcal H$ such that $0\leq E \leq I$.
For a single event, the generalized probability measures on the set of effects are functions $E\mapsto v(E)$ with the properties
\begin{eqnarray}
\label{P1}&& 0\leq v(E) \leq 1\,,\\
\label{P2}&& v(I)=1\,,\\
\label{P3}&& v(E+F+\cdots)=v(E)+v(F)+\cdots
\end{eqnarray}
where $E+F+\cdots \leq I$. A theorem~\cite{bush03} states that, if Eqs.~\eqref{P1}-\eqref{P3} are satisfied, then the probability corresponding to the event represented by $E$ is $v(E)=\Tr{E \rho}$ for some density matrix $\rho$.
For two events, we can define a function $v(E,F)$ with the properties~\cite{Francica222}
\begin{eqnarray}
\label{Q1}&& v(E,F)\in \R \,,\\
\label{Q2}&& v(I,E)=v(E,I)=v(E)\,,\\
\nonumber && v(E+F+\cdots,G)=v(E,G)+v(F,G)+\cdots\,,\\
\label{Q3}&& v(G,E+F+\cdots)=v(G,E)+v(G,F)+\cdots
\end{eqnarray}
where $E+F+\cdots \leq I$.
If Eqs.~\eqref{Q1}-\eqref{Q3} are satisfied, and if $v(E,F)$ is sequentially continuous in its arguments, then the quasiprobability corresponding to the events $E\land F$ is a bilinear function, in detail it is $v(E,F)=\text{Re}\Tr{E F \rho}$ for some density matrix $\rho$ (see Ref.~\cite{Francica222} for details).
Similarly, for three events, we define a quasiprobability $v(E,F,G)$ with the properties
\begin{eqnarray}
\label{Qq1}&& v(E,F,G)\in \R \,,\\
\label{Qq2}&& v(I,E,F)=v(E,I,F)=v(E,F,I)=v(E,F)\,,\\
\nonumber && v(E+F+\cdots,G,H)=v(E,G,H)+v(F,G,H)+\cdots\,,\\
\nonumber && v(G,E+F+\cdots,H)=v(G,E,H)+v(G,F,H)+\cdots\,,\\
\label{Qq3}&& v(G,H,E+F+\cdots)=v(G,H,E)+v(G,H,F)+\cdots\,,
\end{eqnarray}
and in general, for an arbitrary number of events, we define a quasiprobability $v(E,F,\cdots)$ with the properties
\begin{eqnarray}
\label{Q1M}&& v(E,F,\cdots)\in \R \,,\\
\label{Q2M}&& v(I,E,F,\cdots)=v(E,I,F,\cdots)=\cdots=v(E,F,\cdots)\,,\\
\nonumber && v(E+F+\cdots,G,\cdots)=v(E,G,\cdots)+v(F,G,\cdots)+\cdots\,,\\
\label{Q3M}&& \cdots
\end{eqnarray}
where $E+F+\cdots \leq I$. Analogously, if Eqs.~\eqref{Q1M}-\eqref{Q3M} are satisfied, and if $v(E,F,\cdots)$ is sequentially continuous in its arguments, then the joint quasiprobability corresponding to the events $E\land F\land \cdots$ can be expressed as an arbitrary affine combination of $\text{Re} \Tr{X_i \rho}$ where the operators $X_i$ are all the possible products of the effects, e.g., for two events we can consider only the product $X_1=EF$, since $X_2=FE$ gives the same quasiprobability; for three events, we can consider the three products $X_1=EFG$, $X_2=FEG$ and $X_3=EGF$, and so on.
Basically, the quasiprobability is not fixed for more than two events since the proposition $E\land F\land \cdots$ is not well defined.
In particular, for more than two events, the quasiprobability depends on how the events are grouped together.
For instance, for three events, a proposition $E\land F \land G$ can be decomposed in three different ways, which are $E\land F$, $F \land G$, or $F\land E$, $ E\land G$, or $E\land G$, $ G\land F$, then there is a one to one correspondence between the quasiprobabilities $\text{Re} \Tr{X_i \rho}$ and the different decompositions. It is straightforward to see that this correspondence holds also for an arbitrary number of events, thus we can associate the quasiprobability  $v(E,F,G,\ldots)=\text{Re} \Tr{EFG\cdots \rho}$ to the decomposition $E\land F$, $F\land G$, $G\land \cdots$.

\section{Time reversed quasiprobability distribution of work}

We proceed by focusing on the statistics of the work performed in a out-of-equilibrium process. We consider as usual a quantum coherent process generated through a time-dependent Hamiltonian $H(t)=\sum \epsilon_k(t) \ket{\epsilon_k(t)}\bra{\epsilon_k(t)}$ where $\ket{\epsilon_k(t)}$ is the eigenstate with eigenvalue $\epsilon_k(t)$ at the time $t$. The time evolution operator is $U_{t,0}=\mathcal T e^{-i\int_0^t H(s) ds}$, where $\mathcal T$ is the time order operator, and the initial density matrix is $\rho$. We introduce the projectors $\Pi_i = \ket{\epsilon_i}\bra{\epsilon_i}$ and $\Pi'_k=U_{\tau,0}^\dagger \ket{\epsilon'_k}\bra{\epsilon'_k} U_{\tau,0}$, which will represent our events, where $\epsilon_i=\epsilon_i(0)$ and $\epsilon'_k=\epsilon_k(\tau)$.
In general, the work will be represented in terms of the events $\Pi_{i}$, $\Pi_j$, $\cdots$, $\Pi'_k$, $\cdots$, thus a quasiprobability distribution has the form~\cite{Francica222}
\begin{equation}\label{eq. p}
p(w) = \sum_{i,j,\ldots} v(\Pi_i,\Pi_j ,\ldots, \Pi'_k,\ldots ) \delta(w-w(\epsilon_i,\ldots))\,,
\end{equation}
where the quasiprobability $v(\Pi_i,\Pi_j, \ldots, \Pi'_k,\ldots )$ is linear in the initial density matrix $\rho$, e.g., $v(\Pi_i,\Pi'_k ) = \text{Re}\Tr{\Pi_i\Pi'_k\rho}$ if there are only two events, which are $\Pi_i$ and $\Pi'_k$. On the other hand, the support is given by $w(\epsilon_i,\ldots)$ which in general is a function of the eigenvalues of the initial and final Hamiltonian.

For a given representation with quasiprobability distribution $p(w)$, what is the corresponding time reversed representation?
To give an answer, we note that in the time reversed process, the initial state will be $\bar{\rho}$, and the time evolution operator $\bar{U}_{\tau,0}=U_{\tau,0}^\dagger$. Furthermore, the events will be $\bar{\Pi}_i$ and $\bar{\Pi}'_k$ and the work $\bar{w}(\epsilon_i,\ldots)$.
Then, a quasiprobability distribution for the backward process will have the form
\begin{equation}\label{eq. tr}
\bar{p}(w) = \sum_{i,j,\ldots} \bar{v}(\bar{\Pi}_i,\bar{\Pi}_j ,\ldots, \bar{\Pi}'_k,\ldots ) \delta(w-\bar{w}(\epsilon_i,\ldots))\,,
\end{equation}
where $\bar{v}(\bar{\Pi_i},\bar{\Pi}_j ,\ldots, \bar{\Pi}'_k,\ldots )$ is calculated with respect to the initial density matrix $\bar{\rho}$, e.g., $\bar{v}(\bar{\Pi}_i,\bar{\Pi}'_k ) = \text{Re}\Tr{\bar{\Pi}_i\bar{\Pi}'_k\bar{\rho}}$ if there are only two events, which are $\bar{\Pi}_i$ and $\bar{\Pi}'_k$.
Given  a quasiprobability distribution $p(w)$ for the forward process, we can get the corresponding time reversed quasiprobability distribution $\bar{p}(w)$ from it by performing the time reversal, i.e., the replacements $\Pi_i \mapsto \bar{\Pi}_i$, $\Pi'_k \mapsto \bar{\Pi}'_k$, $\rho\mapsto \bar{\rho}$ and $w(\epsilon_i,\ldots)\mapsto \bar{w}(\epsilon_i,\ldots)$. This can be understood  by noting that a representation is defined in terms of certain events and initial density matrix, thus to obtain the corresponding time-reversed one, we have to calculate the time-reversal of these events and initial density matrix.
A natural choice is  $\bar{\rho}= U_{\tau,0} \rho U_{\tau,0}^\dagger$, $\bar{\Pi}_i=U_{\tau,0}\Pi_i U_{\tau,0}^\dagger$ and $\bar{\Pi}'_k=U_{\tau,0}\Pi'_k U_{\tau,0}^\dagger$, then for the quasiprobabilities we will get $\bar{v}(\bar{\Pi}_i,\bar{\Pi}_j ,\ldots, \bar{\Pi}'_k,\ldots ) =v(\Pi_i,\Pi_j ,\ldots, \Pi'_k ,\ldots) $, i.e., they are invariant under time reversal. To prove it, it is enough to note that the quasiprobability involves the real part of a trace of the product of the projectors and the initial density matrix.
Furthermore, by requiring that the work is odd under time reversal, we have $\bar{w}(\epsilon_i,\ldots)=-w(\epsilon_i,\ldots)$, from which we get the time-reversal symmetry relation for work
\begin{equation}\label{eq. sym}
\bar{p}(w) = p(-w)\,.
\end{equation}
For instance, let us focus on the class of quasiprobability distributions
\begin{equation}\label{eq. pq}
p_q(w) = \sum_{i,j,k} \text{Re}\Tr{\Pi_i \rho \Pi_j \Pi'_k} \delta(w-\epsilon'_k + q \epsilon_i + (1-q)\epsilon_j)\,,
\end{equation}
which reproduce the two-projective measurements scheme when the initial state $\rho$ is incoherent with respect to the projectors $\Pi_i$, i.e., $p_q(w)=p_{\text{TPM}}(w)$ when $\rho=\Delta(\rho)$ for any $q$, with
\begin{equation}\label{eq. TPM}
p_{\text{TPM}}(w) = \sum_{i,k} \Tr{\Pi_i \rho \Pi_i \Pi'_k} \delta(w-\epsilon'_k + \epsilon_i)\,,
\end{equation}
and $\Delta(\rho) = \sum_i \Pi_i\rho\Pi_i$. Of course the probability $\Tr{\Pi_i \rho \Pi_i \Pi'_k}$ in Eq.~\eqref{eq. TPM} is of the form $v_{TPM}(E,F)=\Tr{E F E \rho}$, so that does not satisfy Eq.~\eqref{Q3}.

The time reversal of the quasiprobability distribution $p_q(w)$ reads
\begin{equation}
\bar{p}_q(w) = \sum_{i,j,k} \text{Re}\Tr{\bar{\Pi}_i \bar{\rho} \bar{\Pi}_j \bar{\Pi}'_k} \delta(w- q \epsilon_i - (1-q)\epsilon_j+\epsilon'_k)\,,
\end{equation}
which reproduces the two-projective measurements scheme when the initial state $\bar{\rho}$ is incoherent with respect to the projectors $\bar{\Pi}'_k$ for $q=0,1$ but not in general, i.e., $\bar{p}_q(w)=\bar{p}_{\text{TPM}}(w)$ when $\bar{\rho}=\bar{\Delta}(\bar{\rho})$ for $q=0,1$ but not for every $q$, with
\begin{equation}
\bar{p}_{\text{TPM}}(w) = \sum_{i,k} \Tr{\bar{\Pi}'_k \bar{\rho} \bar{\Pi}'_k \bar{\Pi}_i} \delta(w-\epsilon_i+\epsilon'_k)\,,
\end{equation}
and $\bar{\Delta}(\bar{\rho}) = \sum_k \bar{\Pi}'_k\bar{\rho}\bar{\Pi}'_k$.
Thus, since in general $\bar{p}_q(w)\neq\bar{p}_{\text{TPM}}(w)$ when $\bar{\rho}=\bar{\Delta}(\bar{\rho})$, the forward class with the quasiprobability distributions of Eq.~\eqref{eq. pq} is not mapped into the same backward class by performing a time reversal. In detail, the backward class has the quasiprobability distributions
\begin{equation}
\tilde{p}_q(w) = \sum_{i,k,l} \text{Re}\Tr{\bar{\Pi}'_k \bar{\rho} \bar{\Pi}'_l \bar{\Pi}_i} \delta(w- \epsilon_i+q \epsilon'_k + (1-q)\epsilon'_l)\,,
\end{equation}
such that $\tilde{p}_q(w)=\bar{p}_{\text{TPM}}(w)$ when $\bar{\rho}=\bar{\Delta}(\bar{\rho})$. Thus,  we get
\begin{equation}
\tilde{p}_q(w) = \bar{p}_q(w) = p_q(-w)
\end{equation}
for $q=0,1$, but in general $\tilde{p}_q(w) \neq\bar{p}_q(w)= p_q(-w)$ for $q\neq 0,1$. 

In the end, we note that $\tilde{p}_q(w)$ will come from the time-reversal of an operational procedure. We recall that, as shown in Ref.~\cite{Francica22},  the characteristic function $\chi_q(u)=\int e^{iuw}p_q(w)dw $ can be obtained  by introducing a detector in the initial state $\rho^D_0$. The time evolution of the total system is generated by $H(t) - \alpha(t) \Lambda \otimes H(t)$ where $\Lambda$ is a detector observable and $\alpha(t) = \delta(t-\tau+0^+)-\delta(t-0^+)$. When the total system is prepared in the initial state $\rho^D_0\otimes \rho$, the coherence of the detector state can be expressed as
\begin{equation}
\frac{\bra{\lambda}\rho^D_\tau\ket{\lambda'}}{\bra{\lambda}\rho^D_0\ket{\lambda'}} = \Tr{e^{-i\lambda H(0)}\rho e^{i \lambda' H(0)} e^{i(\lambda -\lambda') H^{(H)}(\tau)}}\,,
\end{equation}
where $H^{(H)}(t) = U_{t,0}^\dagger H(t) U_{t,0}$, $\ket{\lambda}$ is the eigenstate of $\Lambda$ with eigenvalue $\lambda$ and $\rho^D_\tau$ is the time evolved detector state. Thus we have
\begin{equation}
\chi_q(u) = \frac{1}{2}\left(\frac{\bra{uq}\rho^D_\tau\ket{u(q-1)}}{\bra{uq}\rho^D_0\ket{u(q-1)}}+ \frac{\bra{u(1-q)}\rho^D_\tau\ket{-uq}}{\bra{u(1-q)}\rho^D_0\ket{-uq}}\right)\,.
\end{equation}
On the other hand, if we realize a time-reversed version of this protocol,  we will get the characteristic function of $\tilde{p}_q(w)$. Thus, the quasiprobability distribution $\tilde{p}_q(w)$ is a different notion of time reversed quasiprobability distribution which is defined operationally. In general, the two notions of time reversed quasiprobability distribution are not compatible if $q\neq 0,1$.

\section{Conclusions}
In this short note we discussed the time-reversal of a general quasiprobability distribution of work. We have shown that an opportune definition of time reversed quasiprobability distribution still gives a time-reversal symmetry relation for work. Furthermore, the time reversal of a quasiprobability that satisfies certain fundamental conditions, such as the reproduction of the two-projective measurements scheme when the initial state is incoherent, in general will not satisfy the same conditions. 

\end{document}